\documentclass[10pt, twocolumn]{article}

\usepackage{authblk}
\usepackage{graphicx}
\usepackage{amsmath}
\usepackage{amsthm}
\usepackage{amssymb}
\usepackage{bbm}
\usepackage{epstopdf}

\usepackage{algorithmic}
\usepackage{algorithm}
\usepackage{hyperref}

\theoremstyle{plain}

\theoremstyle{definition}

\title{Maximum Coverage and Maximum Connected Covering in Social Networks with Partial Topology Information}
\author[1]{Patricio Reyes
\thanks{Email: \href{mailto:preyes@est-econ.uc3m.es}{preyes@est-econ.uc3m.es}}}
\affil[1]{
Dept. Statistics\\
Universidad Carlos III de Madrid (UC3M)\\
Calle Madrid 126, UC3M - Estad\'istica\\
28903 Getafe, Madrid, Spain}
\author[2]{Alonso Silva
\thanks{Email: \href{mailto:alonso.silva@alcatel-lucent.com}{alonso.silva@alcatel-lucent.com}}}
\affil[2]{
Alcatel-Lucent Bell Labs France\\
Centre de Villarceaux\\
Route de Villejust\\
91620 Nozay\\ France}

\date{}

\begin{document}
\maketitle

\begin{abstract}
Viral marketing campaigns seek to recruit the most influential individuals to cover the largest target audience. This can be modeled as the well-studied maximum coverage problem. 
There is a related problem when the recruited nodes are connected. It is called the maximum connected cover problem. This problem ensures a strong coordination between the influential nodes which are the backbone of the marketing campaign. In this work, we are interested on both of these problems.
Most of the related literature assumes knowledge about the topology of the network. Even in that case, the problem is known to be NP-hard. In this work, we propose heuristics to the maximum connected cover problem and the maximum coverage problem with different knowledge levels about the topology of the network. We quantify the difference between these heuristics and the local and global greedy algorithms.
\end{abstract}

\section{Introduction}

One of the main objectives of viral marketing campaigns
is to find the most influential individuals to cover the largest
target audience.
This problem can be modeled as
the well-studied maximum coverage problem.
In the need of coordination through the marketing campaign,
a more relevant objective is to seek the most influential connected individuals.
Hereby the connectedness will be fundamental since
the advertisement needs to spread quickly through the network.
In this work,
we are interested on both of these problems.
Most of the related works on
these topics
assume knowledge about the topology
of the network.
Even in that case, the problem is known to be NP-hard.
Recently, in~\cite{Avrachenkov2012}
the authors propose a (local) greedy algorithm
to the maximum connected covering problem
by learning the
topology of the network on-the-fly.

In this work,
we present different heuristics to both of these problems
with different levels of knowledge about the topology of the network.
We quantify the difference between these
algorithms.
Obviously, different knowledge about the topology of the network
will restrict us to use different heuristics with the problem at hand.

Works providing heuristics
to maximize the impact of a virus marketing campaign
are~\cite{DomingosR2001} and \cite{RichardsonD2002}.
Other works have been interested on the spreading
of information through cascades on a weighted influence graph
and some internal conviction threshold of the individuals (see e.g.~\cite{Kempe2003,Kempe2005,MosselR2007}).
Their work is based on the submodularity of the local spreading
and the bounds on the performance of the greedy algorithms for submodular functions given in~\cite{NemhauserWF1978}.
Our work is different since we are not interested on
the spreading of information through cascades,
but on a ``quick'' spreading:
we are interested on a one-hop
spread of information only through neighbors.
Closely related works are~\cite{Avrachenkov2012} and \cite{GuhaK96},
were the authors assume a one-hop lookahead~\cite{Avrachenkov2012}
and two-hops lookaheads~\cite{GuhaK96}.

The paper is structured as follows.
In Section~\ref{sec:problem-formulation},
we formulate both, the maximum coverage problem
and the maximum connected covering problem.
In Section~\ref{sec:information-levels},
we describe the different levels of
knowledge about the topology of the
network that we consider in this work.
In Section~\ref{sec:description-of-algorithms},
we present existing heuristics and we present
some new heuristics to these problems
based on the different levels of knowledge of the network.
In Section~\ref{sec:simulations},
we present the simulations of the algorithms
and finally we conclude in Section~\ref{sec:conclusions}.

\section{Problem Formulation}\label{sec:problem-formulation}

We consider an influence graph $G=(V,E)$
where $V$ is the set of vertices and
$E\subseteq V\times V$ is the set of edges.
Each vertex of the graph represents an individual
and each edge represents a relationship of mutual influence between them
(e.g. friendship over a social network).
An individual $i\in V$ has influence over another individual $j\in V$
if and only if $\{i,j\}\in E$.
We assume that the influence graph $G$ is an undirected graph with no self-loops.
We denote by $\mathcal{N}(i)$ the set of neighbors of vertex $i$,
i.e., $\mathcal{N}(i)=\{j\in V:\,\{i,j\}\in E\}$,
and for a set of vertices $A\subseteq V$,
we denote by $\mathcal{N}(A)$ the set of neighbors of $A$ as
$\mathcal{N}(A)=\{j\in V\setminus A:\,\textrm{exists }i\in A\,\textrm{such that}\,\{i,j\}\in E\}$.

We consider that time is slotted, i.e., $t\in\mathbb{N}\cup\{0\}$.
We denote by $\mathcal{R}(t)$ the set of recruited individuals at time~$t\ge0$.
In particular, $\mathcal{N}(\mathcal{R}(t))$ is the set containing unrecruited neighbors of $\mathcal{R}(t)$.

The algorithms that we present are sequential algorithms which proceed as follows:
at time $t$, with $0\le t\le K$, the algorithm recruits a node $i\in V\setminus\mathcal{R}(t-1)$ 
and performs the update $\mathcal{R}(t)=\mathcal{R}(t-1)\cup\{i\}$.

The objective of the maximum coverage algorithms is to maximize
the size of the network covering
$\mathcal{C}(t)=\mathcal{R}(t)\cup\mathcal{N}(\mathcal{R}(t))$
and in the case of the maximum connected covering problem
this objective is subject to the additional constraint that the set $\mathcal{R}(t)$
must be connected.

The degree $d(i)$ of a node $i\in V$ is the number of neighbors of a node,
 i.e., $d(i)=\lvert\mathcal{N}(i)\rvert$ where $\lvert\cdot\rvert$
is the cardinality function.
The observed degree $d_{obs}(i,t)$ of a node $i\in V$ at time $t$ is
the number of recruited neighbors or neighbors of recruited neighbors of $i$, i.e.,
$d_{obs}(i,t)=
\lvert
\{
j\in\mathcal{R}(t)\cup\mathcal{N}(\mathcal{R}(t))
:\, \{i,j\}\in E
\}
\rvert$.
The excess degree $d_{excess}(i,t)$ of a node $i\in V$ at time $t$
is difference between the degree and the observed degree of node $i$ at time $t$, i.e.,
$d_{excess}(i,t)=d(i)-d_{obs}(i,t)$.

\section{Information Levels}\label{sec:information-levels}

For both of the problems we are dealing with in this work,
we consider different levels of information about the
topology of the network.
\begin{enumerate}
\item\label{item:list-of-nodes} List of nodes:
we consider that the recruiter knows the list of nodes (we know $V$)
so there is a knowledge about the nodes
the network has and there is a possibility to recruit
any node within the network.
Once a node has been recruited we consider that
the recruited node gives information
about who are its neighbors.
\item\label{item:one-hop-lookahead} One-hop lookahead:
we consider that the recruiter knows only one node,
denoted $i$, and once a node is recruited
it gives information about who are its neighbors
and who are their mutual neighbors (between
recruited nodes).
Actually, the recruiter may only need to know
the quantity of neighbors, observed neighbors
and mutual neighbors (between recruited nodes),
in order to compute the excess degree.
\item\label{item:two-hops-lookahead} Two-hops lookahead:
we consider that the recruiter knows only one node, denoted $i$,
and once a node is recruited
it gives information about who are its neighbors
and neighbors of neighbors and who are
their neighbors and the mutual neighbors.
\item List of nodes and two-hops lookahead:
we have knowledge about the list of nodes
as in \ref{item:list-of-nodes})
and two-hops lookahead as in \ref{item:two-hops-lookahead}).
\item\label{item:full-knowledge} Full knowledge:
we consider that the recruiter has full knowledge
about the topology of the network.
It knows the set of nodes $V$ and the set of edges~$E$.
\end{enumerate}

We notice that in~\cite{Avrachenkov2012}
the knowledge level that they consider is \ref{item:one-hop-lookahead})
since in their case, they do not have any information about
the network topology and they are discovering the network
while they are recruiting over the network.

\section{Description of Algorithms}\label{sec:description-of-algorithms}

In this section, we give a brief description of the algorithms
for the different scenarios (levels of information)
in both problems: the maximum coverage problem (SCP)
and maximum connected covering (MCC) problem.

\subsection{Set Covering Problem (SCP)}

In the first scenario,
called SCP~\ref{algoscp1},
we consider that the recruiter knows the list
of nodes but doesn't have
any information about the topology of the graph
as in \ref{sec:information-levels} \ref{item:list-of-nodes}).
Once we recruit a node and only then,
we consider that the node gives us information
about which nodes it is connected to.
Under these characteristics,
we consider Algorithm~\ref{algoscp1}.
Given that initially you don't have
any information about the topology
of the network, Algorithm~\ref{algoscp1}
simply chooses to recruit a node at random
and since then the recruiter knows to which nodes
it is connected to,
it can remove those nodes (since they are already
covered) from the uncovered nodes and
then again choose a node from within
the set of remaining uncovered nodes
at random.

For the probability distribution over a set of nodes $S\subseteq V$,
we identify each node $i\in S$ with a unique integer from $1$ to $\lvert S\rvert$.
We consider a probability distribution $\zeta$ over the set of nodes $\lvert S\rvert$,
i.e., $\zeta(i)\ge0$ and \mbox{$\sum_{i\in S}\zeta(i)=1$}.
For simplicity, we consider the two following cases:
\begin{itemize}
\item The uniform distribution $\zeta_1(i)=1/\lvert S\rvert$,
\item The degree distribution $\zeta_2(i)=d(i)/\sum_{j\in V} d(j)$.
\end{itemize}
However, we notice that the probability distribution~$\zeta$ is not
restricted to these two choices.

\begin{algorithm}
\caption{SCP 1: Random}
\begin{algorithmic}[1]
\STATE Initialize the list of uncovered nodes $U$ with the set of all nodes $U\gets V$,
the list of recruited nodes $R$ with the empty set $R\gets\emptyset$,
and the list of covered nodes $C$ with the empty set $C\gets\emptyset$,
\STATE $k\gets1$,
\REPEAT
\STATE Recruit a node $i\in U$ uniformly at random, i.e., $R\gets R\cup\{i\}$,
\STATE Remove node $i$ and its neighbors $\mathcal{N}(i)$
from the list of uncovered nodes, i.e., \mbox{$U\gets U\setminus (i\cup\mathcal{N}(i))$}
\STATE Add node $i$ and its neighbors $\mathcal{N}(i)$ to the list of covered nodes,
i.e., \mbox{$C\gets C\cup(i\cup\mathcal{N}(i))$}
\STATE $k\gets k+1$,
\UNTIL{$k>K$ or $U\gets \emptyset$}
\end{algorithmic}
\label{algoscp1}
\end{algorithm}

In the second scenario, called SCP~\ref{algoscp2},
we assume that when a node is recruited
it provides a two-hops lookahead information, i.e.,
it gives information about its neighbors and
the neighbors of its neighbors as in~\ref{sec:information-levels} \ref{item:two-hops-lookahead}).
To take advantage of this knowledge, Algoritm~\ref{algoscp2}
which was originally proposed by Guha and Khuller~\cite{GuhaK96},
proposes to recruit the node from within a two-hop neighborhood
that have the maximum number of uncovered neighbors (maximize the excess degree),
and then again to choose the node from within a two-hop
neighborhood of the set of recruited nodes
that have the maximum number of uncovered neighbors.

\begin{algorithm}
\caption{SCP 2: Two-hops Greedy Algorithm\cite{GuhaK96}}
\begin{algorithmic}[1]
\STATE Initialize the list of uncovered nodes $U$ with the set of all nodes $U\gets V$,
the list of recruited nodes $R$ with the empty set $R\gets\emptyset$,
and the list of covered nodes $C$ with the empty set $C\gets\emptyset$,
\STATE $k\gets1$,
\REPEAT
\STATE Recruit a node $i\in U\cap[\mathcal{N}(R)\cup\mathcal{N}(\mathcal{N}(R))]$
of maximum excess degree, i.e., $R\gets R\cup\{i\}$ where
$i$ is such that $\lvert \mathcal{N}(i)\setminus(R\cup\mathcal{N}(R))\rvert$ is maximum
restricted to the set \mbox{$U\cap[\mathcal{N}(R)\cup\mathcal{N}(\mathcal{N}(R))]$},
\STATE Remove node $i$ and its neighbors $\mathcal{N}(i)$
from the list of uncovered nodes, i.e., \mbox{$U\gets U\setminus (i\cup\mathcal{N}(i))$},
\STATE Add node $i$ and its neighbors $\mathcal{N}(i)$ to the list of covered nodes,
i.e., \mbox{$C\gets C\cup(i\cup\mathcal{N}(i))$}
\STATE $k\gets k+1$
\UNTIL{$k>K$ or $U\gets\emptyset$}
\end{algorithmic}
\label{algoscp2}
\end{algorithm}

In the third scenario, called SCP~\ref{algoscp3},
we consider that the recruiter
knows the list of nodes (as in the first scenario)
and that when a node is recruited
it provides a two-hop lookahead information (as in the second scenario).
To take advantage of this knowledge,
we propose Algorithm~\ref{algoscp3} that at every step
with probability $\delta$ recruits
a node at random from within the set of uncovered nodes
and with probability $(1-\delta)$ recruits
the node from within a two-hop neighborhood
that have the maximum number of uncovered neighbors.
It is clear that the appeal from this version of the algorithm
is that it is a probabilistic combination from both previous scenarios.

For the probability distribution over a set of nodes $S\subseteq V$,
we consider $\zeta$ as in the first scenario.
We consider $\alpha$ to be a variable to be chosen $0\le\alpha\le1$.

\begin{algorithm}
\caption{SCP 3: THG + Random $\alpha$}
\begin{algorithmic}[1]
\STATE Initialize the list of uncovered nodes $U$ with the set of all nodes $U\gets V$,
the list of recruited nodes $R$ with the empty set $R\gets\emptyset$,
and the list of covered nodes $C$ with the empty set $C\gets\emptyset$,
\STATE $k\gets1$,
\REPEAT
\STATE Draw a Bernoulli random variable $X$ with parameter~$\alpha$
\IF{$X=1$}
\STATE Recruit a node $j\in U$ at random (according to $\zeta$) from the set~$U$, i.e., $R\gets R\cup\{j\}$
\STATE Remove node $j$ and its neighbors $\mathcal{N}(j)$ from the list of uncovered nodes,
i.e., \mbox{$U\gets U\setminus (j\cup\mathcal{N}(j))$}
\STATE Add node $j$ and its neighbors $\mathcal{N}(j)$ to the list of covered nodes,
i.e., \mbox{$C\gets C\cup(j\cup\mathcal{N}(j))$}
\ELSE
\STATE Recruit a node $i\in U\cap[\mathcal{N}(R)\cup\mathcal{N}(\mathcal{N}(R))]$
of maximum excess degree, i.e., $R\gets R\cup\{i\}$ where
$i$ is such that $\lvert \mathcal{N}(i)\setminus(R\cup\mathcal{N}(R))\rvert$ is maximum
restricted to the set \mbox{$U\cap[\mathcal{N}(R)\cup\mathcal{N}(\mathcal{N}(R))]$},
\STATE Remove node $i$ and its neighbors $\mathcal{N}(i)$ from the list of uncovered nodes, i.e., \mbox{$U\gets U\setminus (i\cup\mathcal{N}(i))$}
\STATE Add node $i$ and its neighbors $\mathcal{N}(i)$ to the list of covered nodes,
i.e., \mbox{$C\gets C\cup(i\cup\mathcal{N}(i))$}
\ENDIF
\STATE $k\gets k+1$
\UNTIL{$k>K$ or $U\gets\emptyset$}
\end{algorithmic}
\label{algoscp3}
\end{algorithm}

The fourth scenario, called SCP~\ref{algoscp4},
is the full knowledge scenario as in ~\ref{sec:information-levels} \ref{item:full-knowledge} where you
know the topology of the network (the list of nodes,
the list of neighbors of the nodes, the list of neighbors of the neighbors of the nodes,
etc). Algorithm~\ref{algoscp4}
chooses at each step greedily
the node that have the maximum number of uncovered neighbors
from the full set of uncovered nodes.

\begin{algorithm}
\caption{SCP 4: Greedy Algorithm~\cite{NemhauserWF1978}}
\begin{algorithmic}[1]
\STATE Initialize the list of uncovered nodes $U$ with the set of all nodes $U\gets V$,
the list of recruited nodes $R$ with the empty set $R\gets\emptyset$,
and the list of covered nodes $C$ with the empty set $C\gets\emptyset$,
\STATE $k\gets1$,
\REPEAT
\STATE Recruit a node $i\in U$ that maximizes the excess degree, i.e., $R\gets R\cup\{i\}$,
where $i\in U$ is such that $\lvert \mathcal{N}(i)\setminus(R\cup\mathcal{N}(R))\rvert$ is maximum,
\STATE Remove node $i$ and its neighbors $\mathcal{N}(i)$ from the list of uncovered nodes, i.e., \mbox{$U\gets U\setminus (i\cup\mathcal{N}(i))$},
\STATE Add node $i$ and its neighbors $\mathcal{N}(i)$ to the list of covered nodes,
i.e., \mbox{$C\gets C\cup(i\cup\mathcal{N}(i))$}
\STATE $k\gets k+1$,
\UNTIL{$k>K$ or $U\gets \emptyset$}
\end{algorithmic}
\label{algoscp4}
\end{algorithm}

\subsection{Maximum Connected Coverage (MCC) problem}

In the first scenario, called MCC $1$,
we consider that we know a node, denoted node~$i\in V$,
and we consider that when a node is recruited
it gives a one-hop lookahead 
as in \ref{sec:information-levels} \ref{item:one-hop-lookahead}).
In Algorithm~\ref{algomcc1},
we propose a random selection
over the set of neighbors of the recruited nodes
which are not themselves already recruited,
i.e., $P=\mathcal{N}(R)\setminus R$.
We notice that this scenario is different
from a random walk since we are choosing
among the whole set $P$ and not only
the neighbors of the newly recruited node.

In the second scenario, called MCC $2$,
we also consider that we know a node, denoted node~$i\in V$,
and we consider that when a node is recruited
it gives the list of neighbors of the recruited nodes.
In Algorithm~\ref{algomcc3},
which was originally proposed by~\cite{Avrachenkov2012},
the algorithm greedily recrutes
the node in $P$ which maximizes the excess degree.

\begin{algorithm}
\caption{MCC 1: Random Neighbor}
\begin{algorithmic}[1]
\STATE Initialize the list of uncovered nodes $U$ with the set of all nodes $U\gets V$,
the list of recruited nodes $R$ with the empty set $R\gets\emptyset$,
and the list of covered nodes $R$ with the empty set $R\gets\emptyset$,
\STATE Recruit a node $i\in U$ at random (according to $\zeta$), i.e., $R\gets R\cup\{i\}$,
\STATE Remove node $i$ and its neighbors $\mathcal{N}(i)$ from the list of uncovered nodes, i.e., \mbox{$U\gets U\setminus (i\cup\mathcal{N}(i))$},
\STATE Add node $i$ and its neighbors $\mathcal{N}(i)$ to the list of covered nodes,
i.e., \mbox{$C\gets C\cup(i\cup\mathcal{N}(i))$},
\STATE Initialize the list of candidates to be recruited with the set of neighbors of $i$, i.e., $P\gets \mathcal{N}(i)$,
\STATE $k\gets2$
\REPEAT
\STATE Recruit a node $j\in P$ uniformly at random from the set~$P$,
i.e., $R\gets R\cup\{j\}$ with $j\in P$,
\STATE Remove node $j$ from the list of candidates to be recruited,
i.e., \mbox{$P\gets P\setminus\{j\}$},
\STATE Remove the node $j$ and its neighbors $\mathcal{N}(j)$ from the list of uncovered nodes, i.e., \mbox{$U\gets U\setminus (j\cup\mathcal{N}(j))$},
\STATE Add node $j$ and its neighbors $\mathcal{N}(j)$ to the list of covered nodes,
i.e., \mbox{$C\gets C\cup(j\cup\mathcal{N}(j))$},
\STATE Add the unrecruited neighbors of $j$ to the list of candidates to be recruited, i.e.,
\mbox{$P\gets P\cup(\mathcal{N}(j)\cap U)$},
\STATE $k\gets k+1$
\UNTIL{$k>K$ or $U\gets \emptyset$}
\end{algorithmic}
\label{algomcc1}
\end{algorithm}

\begin{algorithm}
\caption{MCC 2: Online Myopic MCC~\cite{Avrachenkov2012}}
\begin{algorithmic}[1]
\STATE Initialize the list of uncovered nodes $U$ with the set of all nodes $U\gets V$,
the list of recruited nodes $R$ with the empty set $R\gets\emptyset$,
and the list of covered nodes $R$ with the empty set $R\gets\emptyset$,
\STATE Recruit node $i\in U$, i.e., $R\gets R\cup\{i\}$,
\STATE Remove node $i$ and its neighbors $\mathcal{N}(i)$
from the list of uncovered nodes, i.e., $U\gets U\setminus(i\cup\mathcal{N}(i))$,
\STATE Add node $i$ and its neighbors $\mathcal{N}(i)$ to the list of covered nodes,
i.e., \mbox{$C\gets C\cup(i\cup\mathcal{N}(i))$},
\STATE $k\gets2$
\REPEAT
\STATE Recruit a node $i\in U$ that maximizes the excess degree, i.e., $R\gets R\cup\{i\}$,
where $i\in U$ is such that $\lvert \mathcal{N}(i)\setminus(R\cup\mathcal{N}(R))\rvert$ is maximum,
\STATE Activate a node $i\in U$ that maximizes the excess degree, i.e., $R=R\cup\{i\}$,
where $i\in U$ is such that $\lvert \mathcal{N}(R)\cap\mathcal{N}(i)\rvert$ is maximum,
\STATE Activate one of the nodes $i\in U$ of maximum excess degree, i.e., $R=R\cup\{i\}$ where
$i$ is such that $d_i-d^{obs}_i=\max_{k\in\{1,\ldots,n\}} d_k-d^{obs}_k$
where $d^{obs}$ is the observed degree.
\STATE Remove the node $i$ and its neighbors $\mathcal{N}(i)$ from the list of uncovered nodes, i.e., \mbox{$U=U\setminus (i\cup\mathcal{N}(i))$}
\STATE $k\gets k+1$
\UNTIL{$k>K$ or $U=\{\emptyset\}$}
\end{algorithmic}
\label{algomcc3}
\end{algorithm}

\section{Simulations}\label{sec:simulations}

We performed simulations of the previously described algorithms in Erdos-Renyi graphs $G(N, p_N)$ 
where \mbox{$N\in\{50, 100, 150, 200, 250\}$} is the number of nodes in the graph and 
$p_N$ is the probability of two nodes being connected.
We chose $p_N = 2\ln(N)/N$ to ensure connectivity.
We simulated $30$ instances for each graph size.
In order to avoid problems for randomly choosing the initial node, 
we set $3$ initial nodes ($10$ instances each) for each algorithm.
In summary, each algorithm was run $3$ times in each graph in each instance, starting by $3$ different nodes.
Therefore, the figures show the average of the number of recruited nodes (over all the graph instances with same size)
needed to cover the whole graph.
We notice that this corresponds to the case when there is no restriction
over $K$ but only on the number of uncovered nodes.

\begin{figure}[h!]
 \centering
  \includegraphics[width=0.49\textwidth]{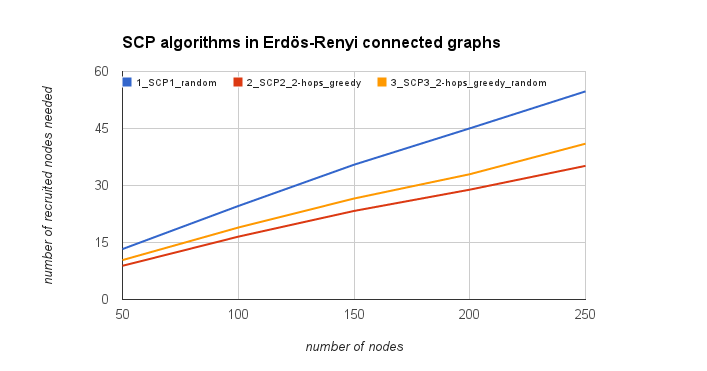}
  \caption{Number of recruited nodes needed vs number of nodes}
  \label{fig:recruited_nodes_vs_nodes1}
\end{figure}

It is difficult to compare different knowledge levels
since for example how can we compare between having 
the possibility of recruiting any node in a network 
but completely at random compared to be able to connect 
only to two hops away nodes but knowing exactly how many observed neighbors and neighbors do they have. 
The first observation that we can make of Figure~\ref{fig:recruited_nodes_vs_nodes1}
is that recruiting nodes at random $\mathrm{SCP}~1$ performs
$56\%$ worst than having a two-hops lookahead and a greedy algorithm $\mathrm{SCP}~3$ ($((\mathrm{SCP}~1-\mathrm{SCP}~3)/\mathrm{SCP}~3)\times100$).
The second observation which we found surprising was 
that in Erdos-Renyi graphs the greedy approach works better 
than the mixed approach (algorithm $\mathrm{SCP}~3$ which combines the 
greedy approach and the random choice). 
The reason why we were expecting to have a different behavior 
is because the algorithm may start in a bad initial location 
and through a greedy approach it may take a while before finding 
good nodes to be recruited. 
In fact, $\mathrm{SCP}~3$ performs $17\%$
worst than the greedy approach $\mathrm{SCP}~3$ ($((\mathrm{SCP}~2-\mathrm{SCP}~3)/\mathrm{SCP}~3)\times100$).
We believe that the performance of $\mathrm{SCP}~3$ may improve 
by modifying the parameter $\alpha$ which we took as $\alpha=1/2$.

We performed simulations of the previously described algorithms also in Barabasi-Albert graphs.
The chosen Barabasi-Albert graphs were undirected graphs and were generated as follows.
We started with a single vertex. 
At each time step, we added one vertex and the new vertex connects two edges to the old vertices. 
The probability that an old vertex is chosen is proportional to its degree.

\begin{figure}[h!]
 \centering
  \includegraphics[width=0.49\textwidth]{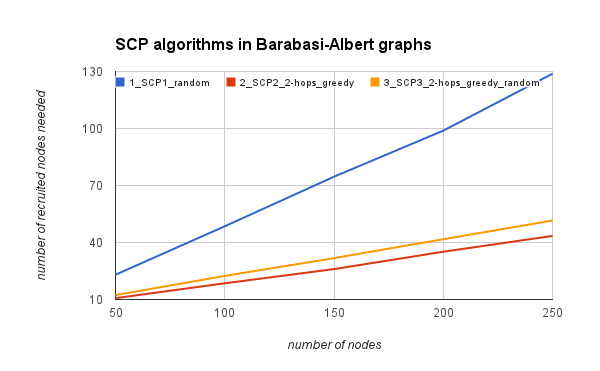}
  \caption{Number of recruited nodes needed vs number of nodes}
  \label{fig:recruited_nodes_vs_nodes3}
\end{figure}

In Figure~\ref{fig:recruited_nodes_vs_nodes3}
(similarly to Figure~\ref{fig:recruited_nodes_vs_nodes1}),
we notice that recruiting nodes at random $\mathrm{SCP}~1$
performs $196\%$ worst than having a two-hops lookahead
and a greedy algorithm $\mathrm{SCP}~3$ ($((\mathrm{SCP}~1-\mathrm{SCP}~3)/\mathrm{SCP}~3)\times100$).
The mixed approach with combines the greedy approach and the random choice $\mathrm{SCP}~2$
performs $19\%$ worst than the greedy approach $\mathrm{SCP}~3$ ($((\mathrm{SCP}~2-\mathrm{SCP}~3)/\mathrm{SCP}~3)\times100$).

\begin{figure}[h!]
 \centering
  \includegraphics[width=0.49\textwidth]{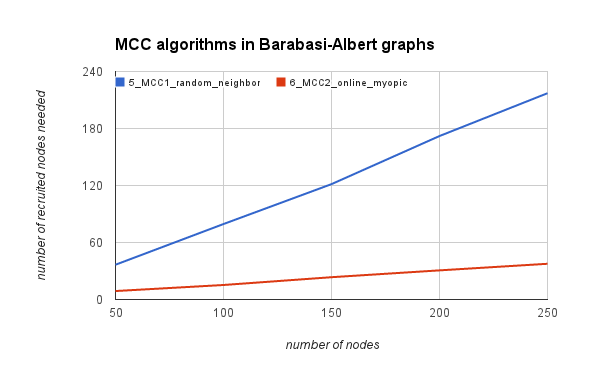}
  \caption{Number of recruited nodes needed vs number of nodes}
  \label{fig:recruited_nodes_vs_nodes4}
\end{figure}

Similarly, in Figure~\ref{fig:recruited_nodes_vs_nodes4},
we have that to choose uniformly at random
between uncovered nodes performs very poorly
compared to the greedy one-hop lookahead algorithm.

\begin{figure}[h!]
 \centering
  \includegraphics[width=0.49\textwidth]{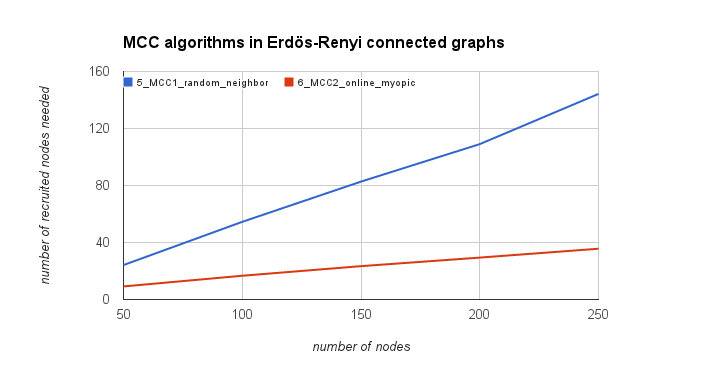}
  \caption{Number of recruited nodes needed vs number of nodes}
  \label{fig:recruited_nodes_vs_nodes2}
\end{figure}

\section{Conclusions and Future Directions}\label{sec:conclusions}

In this work, we were interested on two different problems:
the maximum coverage problem and the maximum connected covering problem.
The motivation of our work is viral marketing campaigns
on social networks.
Our perspective was to analyze both problems from
the knowledge we may have of the topology of the network.
We presented some existing and new heuristics to both of these problems.
We quantified how different levels of information 
have an effect on the type of algorithm
that we choose and this translates on a better or worst
performance depending on the knowledge we have
on the topology of the network.

There are many interesting future directions to this work.
Just to name a few, one direction is to provide theoretical
bounds to the new heuristics and to consider
digraphs instead of undirected graphs.
Another direction is to study how changes on the topology
of the network can affect the problem at hand.
Moreover, if there are changes constantly,
how to make the maximum coverage set and maximum connected covering set
to change together with this dynamicity.

\section*{Acknowledgments}
The work of A.~Silva was partially carried out at LINCS (\url{www.lincs.fr}).

\bibliography{my-bibliography}
\bibliographystyle{hieeetr}
\end{document}